\documentclass{WileyMSP-template}

\usepackage{bm,color,amsmath,txfonts}
\usepackage{graphicx}
\usepackage{siunitx}
\usepackage{subfigure}
\usepackage{verbatim}
\usepackage{dcolumn}% Align table columns on decimal point
\usepackage{bm}% bold math
\usepackage{epsf}
\usepackage{xcolor}
\usepackage{hyperref}
\usepackage{hhline}
\usepackage{float}
\usepackage{enumerate}
\usepackage{bbm}
\usepackage{lipsum}
\usepackage{cite}

\begin{document}

\pagestyle{fancy}
\rhead{\includegraphics[width=2.5cm]{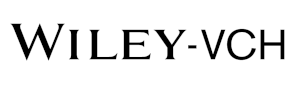}}

\title{Microwave-optics Entanglement via Cavity Optomagnomechanics}

\maketitle

% Author: Please give full first and last names for authors and include * after the name of all corresponding authors

\author{Zhi-Yuan Fan},
\author{Liu Qiu},
\author{Simon Gr\"oblacher},
\author{Jie Li*}

% Dedication

\dedication{}

% Affiliations: Please provide adacemic titles (Prof. or Dr.) for all authors where applicable, and include an institutional email address for all corresponding authors
\begin{affiliations}
Z.-Y. Fan, J. Li\\
Interdisciplinary Center of Quantum Information, State Key Laboratory of Modern Optical Instrumentation,
and Zhejiang Province Key Laboratory of Quantum Technology and Device,
School of Physics\\
Zhejiang University\\
Hangzhou 310027, China\\
Email Address: jieli007@zju.edu.cn

L. Qiu\\
Institute of Science and Technology Austria\\
am Campus 1, 3400, Klosterneuburg, Austria\\

S. Gr\"oblacher\\
Kavli Institute of Nanoscience, Department of Quantum Nanoscience\\
Delft University of Technology\\
2628CJ Delft, The Netherlands\\

\end{affiliations}

% Keywords: Please provide a minimum of three and a maximum of seven keywords, separated by commas

\keywords{cavity magnonics, cavity magnomechanics, optomechanics, microwave-optics entanglement, quantum network}

% Abstract should be written in the present tense and impersonal style (i.e., avoid we), and be at most 200 words long
\begin{abstract}

Microwave-optics entanglement is a vital component for building hybrid quantum networks. Here, a new mechanism for preparing stationary entanglement between microwave and optical cavity fields in a cavity optomagnomechanical system is proposed. It consists of a magnon mode in a ferrimagnetic crystal that couples directly to a microwave cavity mode via the magnetic dipole interaction, and indirectly to an optical cavity through the deformation displacement of the crystal. The mechanical displacement is induced by the magnetostrictive force and coupled to the optical cavity via radiation pressure. Both the opto- and magnomechanical couplings are dispersive. Magnon-phonon entanglement is created via magnomechanical parametric down-conversion, which is further distributed to optical and microwave photons via simultaneous optomechanical beamsplitter interaction and electromagnonic state-swap interaction, yielding stationary microwave-optics entanglement. The microwave-optics entanglement is robust against thermal noise, which will find broad potential applications in quantum
networks and quantum information processing with hybrid quantum systems.

\end{abstract}

% Text: Please use section headings and subheadings as specified below. For communications, all section headings apart from Experimental Section should be removed
% Please make the first reference to a display item bold: \textbf{Figure 1}
% Do not abbreviate Figure, Equation, etc.; display items are always singular, i.e., Figure 1 and 2.
% Equations are always singular, i.e., Equation 1 and 2, and should be inserted using the {equation} environment, not as graphics
% Please do not use footnotes in the text, additional information can be added to the Reference list.

\section{Introduction}
\justifying

Optical entanglement is a crucial resource and finds a wide range of applications in quantum information science, such as quantum teleportation,\textsuperscript{\cite{Wootters,Bouwmeester,Furusawa}} quantum networks,\textsuperscript{\cite{Kimble,Wehner}} quantum logical operations,\textsuperscript{\cite{Milburn}} quantum metrology,\textsuperscript{\cite{VG}} and fundamental tests of quantum mechanics.\textsuperscript{\cite{Hensen,Giustina,Shalm}} Entangled optical fields can be produced in various different ways, such as by exploiting parametric down-conversion in nonlinear crystals,\textsuperscript{\cite{PK,Kwiat}} four-wave mixing in optical fibers\textsuperscript{\cite{fiber1,fiber2}} and atomic vapors,\textsuperscript{\cite{atom1,atom2}} semiconductor quantum dots,\textsuperscript{\cite{dot1,dot2}} and periodically poled lithium niobate waveguide.\textsuperscript{\cite{Gisin}} In the microwave domain, entangled fields are normally generated using Josephson parametric amplifiers.\textsuperscript{\cite{JPA1,JPA2,JPA3}} They can also be created by reservoir engineering of an atomic beam,\textsuperscript{\cite{DV2007}} and injecting a squeezed vacuum through a microwave beamsplitter.\textsuperscript{\cite{Menzel}} In addition, optical (microwave) fields can be entangled in an optomechanical system by coupling them to a common mechanical resonator via radiation pressure.\textsuperscript{\cite{Tombesi,LTian,YDWang,HXTang,SB1,SB2}} 

Despite the above mentioned approaches, there are much fewer efficient ways to produce microwave-optics entanglement due to large frequency mismatch. This special type of optical entanglement finds particularly important applications in quantum information processing with hybrid quantum systems,\textsuperscript{\cite{Hybrid1,Hybrid2,Hybrid3}}  and hybrid quantum networks with distinct quantum nodes working in the microwave and optical frequency ranges.\textsuperscript{\cite{JiePRX1,JiePRX2,JiePRX3,JiePRX4,JiePRX5}} It can be generated by exploiting optomechanical interactions, where both optical and microwave cavity modes couple to a common mechanical oscillator (forming the optoelectromechanical system\textsuperscript{\cite{Forsch1,Forsch2,Forsch3}}), and by activating the parametric down-conversion and state-swap interactions in the electro- and optomechanical subsystems, respectively.\textsuperscript{\cite{Tombesi,LTian,YDWang,HXTang}} Alternatively, it can also be produced by the parametric down-conversion in directly coupled electro-optic systems.\textsuperscript{\cite{Tsang,Fink19,Salamnogli}}

In this work, we present a new mechanism to prepare stationary microwave-optics entanglement in a cavity optomagnomechanical (OMM) configuration exploiting nonlinear magnetostrictive and radiation-pressure interactions. The OMM system consists of a magnon mode in a ferrimagnetic crystal, e.g., yttrium-iron-garnet (YIG), that is dispersively coupled to a phonon mode of the magnetostrictively induced mechanical vibration,\textsuperscript{\cite{SA,Jie18,JieL20,Davis,Jie22}} and the mechanical motion further couples to an optical cavity via radiation pressure.\textsuperscript{\cite{Fan}} The magnon mode further couples to a microwave cavity mode via the magnetic dipole interaction,\textsuperscript{\cite{S1,S2,S3}} forming the cavity-OMM system. We show that by simultaneously driving the magnon mode (the optical cavity) to activate the magnomechanical parametric down-conversion (optomechanical beamsplitter) interaction, magnon-phonon entanglement is created. The entanglement is further transferred to optical and microwave cavity photons due to the effective optomechanical and electromagnonic state-swap interactions, and consequently, a stationary microwave-optics entangled state is prepared.  {Our system is one of the configurations of hybrid quantum systems based on magnons\textsuperscript{\cite{APE}}, which find many promising applications, such as magnon laser and optomagnonic frequency combs etc\textsuperscript{\cite{x1,x2,x3,x4}}.}

\section{System and Model}
\justifying

We consider a cavity-OMM system, as depicted in Figure 1a,b, which consists of a microwave cavity mode, a magnon mode (e.g., the Kittel mode\textsuperscript{\cite{Kittel}}) in a YIG crystal, a mechanical vibration mode, and an optical cavity mode.  The magnon mode is embodied by the collective motion (spin wave) of a large number of spins in the YIG crystal, and is activated by placing the crystal in a uniform bias magnetic field and applying a microwave drive field with its magnetic component perpendicular to the bias field. The magnon mode couples to a microwave cavity field via the magnetic dipole interaction\textsuperscript{\cite{S1,S2,S3}} by putting the crystal near the maximum magnetic field of the cavity mode. The magnetostrictive interaction of the ferrimagnet leads to a dispersive coupling between magnons and lower-frequency vibration phonons for a relatively large size of the crystal.\textsuperscript{\cite{SA,Jie18,JieL20,Davis,Jie22}} The magnomechanical displacement further couples to an optical cavity via radiation pressure.\textsuperscript{\cite{MA,QST23}}  The magno- and optomechanical dispersive couplings play a key role in cooling the mechanical motion and generating entanglement in the system. The YIG crystal can be either a sphere\textsuperscript{\cite{SA,Davis,Jie22}} or a micro bridge structure.\textsuperscript{\cite{QST23,PRAp}} For the convenience of optical cavity fabrication, we adopt the latter and the optical cavity can be realized by attaching a small highly-reflective mirror pad to the surface of the micro bridge.\textsuperscript{\cite{Fan,SG}} {The attached mirror should be fabricated sufficiently small and light, such that it will not appreciably affect the mechanical properties of the micro bridge, e.g. the mechanical displacement. The deformation displacement can be approximately regarded as uniform in the direction perpendicular to the attached surface (with negligible bending displacement), such that the YIG bridge and the mirror pad can stick together tightly and are integrated into one body, which oscillate approximately with the same frequency.}
Alternatively, one may consider using the `membrane-in-the-middle' approach\textsuperscript{\cite{Harris}} to achieve the optomechanical dispersive coupling.  Besides, the micro bridge can be made small to obtain a relatively large magnomechanical bare coupling. {We, however, do not use thin YIG films, where high-frequency (GHz) phonon modes couple linearly (and strongly) to the magnon mode\textsuperscript{\cite{KAn1,KAn2}}. The linear coupling is not desired for our purpose of producing entanglement.}
{One can also consider using the direct mechanical contact approach to realize the optomagnomechanical coupling.\textsuperscript{\cite{Dong}}}
The red-detuned driven optical cavity is used to cool the lower-frequency mechanical motion, while the activated magnomechanical Stokes scattering (by driving the magnons with a blue-detuned microwave field) is sufficient to create entanglement.  This is a key difference from the entanglement mechanism in ref.\cite{Jie18}, where the mechanical motion is cooled by a red-detuned microwave drive field. 

The Hamiltonian of such a hybrid microwave cavity-OMM system reads\textsuperscript{\cite{Jie18,Fan}}
\begin{align}
	\begin{split}
		H/\hbar &= \! \sum_{j=a,m,c}\!\! \omega_{j} j^\dagger j+\frac{\omega_b}{2} \left(q^2+p^2\right) + H_{\rm dri}/\hbar   \\
		& + g_a \left(a^\dagger m+am^\dagger \right) + g_m m^\dagger mq  - g_c c^\dagger cq ,
	\end{split}
\end{align}
where $j=a,m,c$ ($j^\dagger$) are the annihilation (creation) operators of the microwave cavity mode, the magnon mode, and the optical cavity mode, respectively, satisfying the canonical commutation relation $[j,j^\dagger]=1$. $q$ and $p$ ($[q,p]=i$) denote the dimensionless position and momentum of the mechanical vibration mode, and $\omega_k$ ($k\,\,{=}\,\,a,m,c,b$) are the resonant frequencies of the four modes. $g_a$ is the electromagnonic coupling strength, responsible for the microwave-magnon state swapping, and $g_m$ ($g_c$) is the vacuum magnomechanical (optomechanical) coupling strength, which can be significantly enhanced by driving the magnon mode (optical cavity) with a strong field. The driving Hamiltonian $ H_{\rm dri}/\hbar = i \Omega \big(m^\dagger e^{-i\omega_0 t} -{\rm H.c.} \big)+iE \big(c^\dagger e^{-i\omega_L t} - {\rm H.c.} \big)$, where the Rabi frequency $\Omega =\frac{\sqrt{5} }{4} \gamma \sqrt{N}H_d$\textsuperscript{\cite{Jie18}} describes the coupling strength between the magnon mode and the microwave drive field, with the gyromagnetic ratio $\gamma$, the total spin number $N$ of the ferrimagnet, and the amplitude of the drive magnetic field $H_d$. $E=\sqrt{2\kappa_c P_L/(\hbar\omega_L)}$ denotes the coupling strength between the optical cavity and the driving laser, where $P_L$ ($\omega_L$) is the power (frequency) of the laser field, and $\kappa_c$ is the cavity decay rate.

Including the dissipation and input noise of each mode and working in the interaction picture with respect to $\hbar \omega_0 (a^\dagger a+m^\dagger m)+\hbar \omega_L c^\dagger c$, we obtain the quantum Langevin equations (QLEs) of the system, given by
\begin{align}
	\begin{split}
		\dot{a}=&-i\Delta_a a-\kappa _a a-ig_a m+\sqrt[]{2\kappa_a}a_{in},  \\
		\dot{m}=&-i\Delta_m m-\kappa_m m-ig_a a-ig_m mq+\Omega +\sqrt{2\kappa_m}m_{in}, \\
		\dot{q}=&\ \omega_b p,\,\,  \dot{p}= -\omega_b q-\gamma_b p+g_c c^\dagger c-g_m m^\dagger m+\xi, \\
		\dot{c}=& -i\Delta_c c-\kappa_c c +ig_c cq+E+\sqrt{2\kappa_c}c_{in},	
	\end{split}
\end{align}
where $\Delta_{a(m)}=\omega_{a(m)}-\omega_0$, $\Delta_c=\omega_c-\omega_L$, and $\gamma_b$ and $\kappa_j$ $(j=a,m,c)$ are the dissipation rates of the mechanical, microwave cavity, magnon, and optical cavity modes, respectively. $j_{in}$ are the corresponding input noise operators, which are zero-mean and obey the following correlation functions: $ \langle j_{in}(t)j^\dagger _{in}(t') \rangle=\big[ N_j(\omega_j)+1 \big]\delta(t-t')$, and $\langle j^\dagger _{in}(t)j_{in}(t') \rangle=N_j(\omega_j)\delta(t-t') $.  $\xi(t)$ is a Brownian stochastic force acting on the mechanical oscillator, which is non-Markovian by nature, but can be assumed Markovian for a large mechanical quality factor $Q_b=\omega_b/\gamma_b\gg 1$.\textsuperscript{\cite{Kac1,Kac2}} It then takes a $\delta$-autocorrelation: $\left \langle \xi(t)\xi(t')+\xi(t')\xi(t) \right \rangle/2\simeq\gamma_b \big[2N_b(\omega_b)+1 \big] \delta(t-t')$. The mean thermal excitation number of each mode $N_k(\omega_k)=\big[\mathrm{exp}(\hbar\omega_k/k_BT)-1 \big]^{-1}$($k=j,b$), with $T$ being the environmental temperature.

The creation of microwave-optics entanglement requires sufficiently strong magno- and optomechanical coupling strengths. To this end, we apply strong drive fields onto the magnon and optical cavity modes, respectively, which lead to large coherent amplitudes $\left | \left \langle m \right \rangle  \right |, \left | \left \langle c \right \rangle  \right | \gg 1$ at the steady state. This enables us to linearize the system dynamics around the large average values by neglecting small second-order fluctuation terms.\textsuperscript{\cite{DV07}} We therefore obtain a set of linearized QLEs for the quantum fluctuations of the system, which can be written in the matrix form of
\begin{align}
	\dot{u}(t)=Au(t)+n(t),
\end{align}
where $u(t)=\big[\delta X_a(t),\delta Y_a(t), \delta X_m(t), \delta Y_m(t), \delta q(t), \delta p(t), \delta X_c(t)$, $\delta Y_c(t) \big]^T$ is the vector of the quadrature fluctuations of the system, $n(t)=[\sqrt{2\kappa_a}X_{a}^{in}(t), \sqrt{2\kappa_a}Y_{a}^{in}(t),\sqrt{2\kappa_m}X_{m}^{in}(t), \sqrt{2\kappa_m}Y_{m}^{in}(t),$  $0, \xi(t), \sqrt{2\kappa_c}X_{c}^{in}(t), \sqrt{2\kappa_c}Y_{c}^{in}(t)]^T$ is the vector of the input noises, where the quadratures are defined as $X_j(t)=\frac{1}{\sqrt{2}}( j+ j^\dagger)$ and $Y_j(t)\,{=}\frac{i}{\sqrt{2}}( j^\dagger {-} j)$, and $\delta X_j(t)$ and $\delta Y_j(t)$ ($X_{j}^{in}(t)$ and $Y_{j}^{in}(t)$) represent the corresponding fluctuations (input noises), and the drift matrix $A$ is given by
\begin{align}
	A=\begin{pmatrix}
		-\kappa_a & \Delta_a & 0 & g_a & 0 & 0 & 0 &0 \\
		-\Delta_a & -\kappa_a & -g_a & 0 & 0 & 0 & 0 & 0\\
		0 & g_a & -\kappa_m & \tilde{\Delta}_m & G_m & 0 & 0 & 0\\
		-g_a & 0 & -\tilde{\Delta}_m & -\kappa_m & 0 & 0 & 0 & 0\\
		0 & 0 & 0 & 0 & 0 & \omega_b & 0 & 0\\
		0 & 0 & 0 & -G_m & -\omega_b & -\gamma_b & 0 & -G_c\\
		0 & 0 & 0 & 0 & G_c & 0 & -\kappa_c & \tilde{\Delta}_c\\
		0 & 0 & 0 & 0 & 0 & 0 & - \tilde{\Delta}_c & -\kappa_c
	\end{pmatrix}.
\end{align}
We define the effective magno- and optomechanical coupling strengths as $G_m=-i\sqrt{2}g_m\left\langle m \right\rangle$ and $G_c=i\sqrt{2}g_c\left\langle c \right\rangle$, where the steady-state amplitudes are  
\begin{align}
	\left \langle m \right \rangle =\frac{\Omega(\kappa_a+i\Delta_a)}{g_a^2+(\kappa_m+i\tilde{\Delta}_m)(\kappa_a+i\Delta_a)}, \,\,\,\,  \left \langle c \right \rangle =\frac{E}{\kappa_c+i\tilde{\Delta}_c}, 
\end{align}
and the effective detunings $\tilde{\Delta}_m=\Delta_m+g_m\left \langle q \right \rangle $ and $\tilde{\Delta}_c=\Delta_c-g_c\left \langle q \right \rangle $, including the frequency shifts due to the mechanical displacement, which is jointly induced by the magnetostriction and radiation pressure interactions. The steady-state displacement $\left \langle q \right \rangle =\big( g_c\left | \left \langle c \right \rangle  \right |^2-g_m\left | \left \langle m \right \rangle  \right |^2 \big)/\omega_b$.    Note that the above drift matrix $A$ is derived under the optimal conditions for the microwave-optics entanglement, i.e., $\left | \Delta_a \right |,\left| \tilde{\Delta}_m \right| ,\tilde{\Delta}_c \simeq \omega_b \gg \kappa_j $ (see Figure 1c). This yields approximate expressions of the amplitudes, $\left \langle m \right \rangle \simeq i\Omega \Delta_a/\big(g_a^2-\tilde{\Delta}_m\Delta_a \big)$ and $\left \langle c \right \rangle \simeq -iE/\tilde{\Delta}_c$, which are pure imaginary and thus give rise to approximately real coupling strengths $G_m$ and $G_c$.

%By solving the equations of the classical averages at the steady state, we obtain
%      \begin{align}
	%		\begin{split}
		%			&\left \langle a \right \rangle =\frac{-ig_a \left \langle m \right \rangle }{\kappa _a+i\Delta_a},\ \left \langle c \right \rangle =\frac{E}{\kappa_c+i\tilde{\Delta}_c},\\
		%			&\left \langle p \right \rangle=0,\ \left \langle q \right \rangle =(g_c\left | \left \langle c \right \rangle  \right |^2-g_m\left | \left \langle m \right \rangle  \right |^2)/\omega_b,   \\
		%			&\left \langle m \right \rangle =\frac{\Omega(\kappa_a+i\Delta_a)}{g_a^2+(\kappa_m+i\tilde{\Delta}_m)(\kappa_a+i\Delta_a)},\\
		%		\end{split}
	%	\end{align}

We aim to prepare stationary entanglement between microwave and optical cavity fields. The steady state of the system is a four-mode Gaussian state, because of the linearized dynamics and the Gaussian input noises. A Gaussian state is fully characterized by the first and second moments of the quadrature operators. The first moments can be adjusted by local displacement operations in phase space and such operations do not change quantum correlations like entanglement. The state of interest is therefore characterized by an $8\times8$ covariance matrix (CM) $V$, of which the entries are defined as $V_{ij}= \langle u_i(t) u_j(t')+u_j(t') u_i(t) \rangle/2\ $($i,j=1,2,...,8$). The CM in the steady state can be obtained by solving the Lyapunov equation\textsuperscript{\cite{DV07}}
\begin{align}
	AV+VA^T=-D,
\end{align}
where $D=\mathrm{diag} \,[\, \kappa _a(2N_a+1), \kappa _a(2N_a+1), \kappa _m(2N_m +1),$  $\kappa _m(2N_m+1), \,\,0,\,\, \gamma_b(2N_b+1),\ \kappa _a(2N_c+1),\ \kappa _a(2N_c+1)]$ is the diffusion matrix and defined by $D_{ij}\delta(t-t')= \langle n_i(t) n_j(t')+n_j(t') n_i(t) \rangle /2$.  We adopt the logarithmic negativity\textsuperscript{\cite{LN1,LN2,LN3}} to quantify the microwave-optics entanglement, i.e.,
\begin{align}
	E_{ca} = \mathrm{max}\big[ 0,-\mathrm{ln}(2\eta^-) \big],
\end{align}
where $\eta^-\equiv 2^{-1/2} \big[\Sigma - \big( \Sigma^2 - 4\, \mathrm{det}V_{4} \big) ^{1/2} \big] ^{1/2}$, and $V_{4}$ is the $4\times4$ CM associated with the microwave and optical modes. $V_{4}=\big[V_a,V_{ac};V_{ac}^T,V_c \big]$, with $V_a$, $V_c$, and $V_{ac}$ being the $2\times2$ blocks of $V_{4}$, and $\Sigma\equiv \mathrm{det}V_a+\mathrm{det}V_c-2\mathrm{det}V_{ac} $. {Similarly, we can calculate the magnon-optics entanglement ($E_{cm}$) and the microwave-mechanics entanglement ($E_{ab}$).}

\section{Results and Discussions}
\justifying

The mechanism of producing microwave-optics entanglement is as follows. Firstly, the highly thermally populated mechanical mode should be cooled (close) to its quantum ground state in order to generate entanglement in the system. This can be achieved by driving the optical cavity with a red-detuned laser field ($\tilde{\Delta}_c \simeq \omega_b$) to activate the optomechanical anti-Stokes scattering, optimally in the resolved sideband limit $\omega_b \gg \kappa_c$ (Figure 1c).\textsuperscript{\cite{MA}} The effective optomechanical coupling is then a beamsplitter type, which takes the same form as the electromagnonic coupling. As is known, such beamsplitter couplings solely will not create any entanglement, and the system requires a parametric down-conversion process to generate entanglement. This can be realized by driving the magnon mode with a blue-detuned microwave field ($\tilde{\Delta}_m \simeq -\omega_b$) to activate the magnomechanical Stokes scattering, which yields magnon-phonon entanglement.  Further due to the effective beamsplitter (state-swap) interaction in the optomechanical and electromagnonic couplings, the optical and microwave cavity photons therefore get entangled.

The optimal detunings $\tilde{\Delta}_c \simeq - \tilde{\Delta}_m \simeq  \omega_b$ adopted in the tripartite OMM system are analogous to those used to produce optical entanglement in a tripartite optomechanical system, where two cavity fields, either optical or microwave, couple to a common mechanical resonator.\textsuperscript{\cite{Tombesi,LTian,YDWang,HXTang}} The optical entanglement is the result of the combination of simultaneous parametric down-conversion and state-swap interactions associated with the two cavity fields and the mechanical oscillator being an intermediary to distribute the entanglement. This is the fundamental mechanism of producing optical entanglement in optomechanical systems.\textsuperscript{\cite{Tombesi,LTian,YDWang,HXTang,Genes,Gut}}  Due to their similar Hamiltonians and scattering processes, magnons and optical photons are expected to be entangled in our OMM system. This is confirmed by Figure 2a. %The magnon-optics entanglement ($E_{cm}$) is transferred from the magnon-phonon entanglement ($E_{mb}$) via the optomechanical beamsplitter coupling.
Further, the microwave-magnon state-swap interaction enables the distribution of the entanglement to the microwave photons, thus establishing the microwave-optics entanglement, as shown in Figure 2b. The complementary distribution of the entanglement in Figure 2a,b is a clear sign of the entanglement transfer from the magnon-optics subsystem ($E_{cm}$) to the microwave-optics subsystem ($E_{ca}$).   Such entanglement transfer is optimally achieved in the strong-coupling regime, where the microwave-magnon coupling rate is larger than their dissipation rates, $g_a > \kappa_a, \kappa_m$,\textsuperscript{\cite{Jie19a1,Jie19a2}} see Figure 2c. The microwave-optics entanglement $E_{ca}$ achieves its maximum around $g_a/2\pi \simeq 4$ MHz and starts to decline as $g_a$ continues to increase. This is because the entanglement is further distributed to the microwave-mechanics subsystem (Figure 2c, the dot-dashed line). The distribution of entanglement among different subsystems is a typical feature of the entanglement in multipartite systems\textsuperscript{\cite{Jie18}} where entanglement is considered as a finite quantum resource.

Figure 2 is plotted with experimentally feasible parameters\textsuperscript{\cite{SA,Jie18,JieL20,Davis,Jie22,PRAp}}: $\omega_a/2\pi=10$ GHz, $\omega_b/2\pi=40$ MHz, an optical wavelength $\lambda_c=1550$ nm, $\kappa_{a(m)}/2\pi=1.5$ MHz, $\kappa_c/2\pi=2$ MHz, $\gamma_b/2\pi=10^2$ Hz, $g_a/2\pi=4$ MHz, $G_m/2\pi=2$ MHz, $G_c/2\pi=8$ MHz, $T=10$ mK, and the magnon frequency is tuned to be resonant with the microwave cavity ($\Delta_a = \tilde{\Delta}_m$). All the entanglements are in the steady state confirmed by the negative eigenvalues (real parts) of the drift matrix. Under these parameters and the optimal detunings, the effective mean phonon number is $\bar{n}_b^{\rm eff}\simeq 0.17$, indicating that the mechanical mode is significantly cooled to its ground state.  The microwave-optics entanglement is robust against thermal noises, and is present at an environmental temperature up to $T \simeq 380$ mK (Figure 2d). Note that a relatively large optomechanical coupling $G_c/2\pi=8$ MHz is used for cooling, under which the system remains stable due to the multi dissipation channels in our quadripartite system.  This corresponds to a laser power $P_L \simeq 2.56$ mW for $g_c/2\pi=2$ kHz.  The magnomechanical coupling $G_m/2\pi=2$ MHz corresponds to a microwave drive power $P_0 \simeq 0.91$ mW {(the drive magnetic field $H_d \simeq 8.7 \times 10^{-4}$ T)} for a 5 $\times$ 2 $\times$ 1 $\mu$m$^3$ YIG cuboid with $g_m/2\pi=20$ Hz. 
{Note that a much stronger $g_m$ is used than $\sim$10 mHz for a large-size YIG sphere with the diameter in the $10^2$ $\mu$m range,\textsuperscript{\cite{SA,Davis,Jie22}} because we adopt a much smaller micron-sized YIG crystal, which permits a much stronger yet dispersive magnon-phonon coupling.}
To determine the power, we have used the relation of the drive magnetic field $H_d$ and the power $P_0$, i.e., $H_d=\sqrt{2\mu_0 P_0/(lwc)}$, where $\mu_0$ is the vacuum magnetic permeability, $c$ is the speed of the electromagnetic wave propagating in vacuum, and $l$ and $w$ are the length and width of the YIG crystal. 
{This is derived as follows. The time average of energy density is $\rho_e=H_d^2/(2\mu_0)$ and thus the microwave drive power $P_0=\rho_e Ac$, where $A=l \times w$ is the cross-sectional area of the YIG cuboid. The drive magnetic field is applied in the direction perpendicular to the cross-section. This then yields the relation between the drive magnetic field $H_d$ and the power $P_0$, i.e., $H_d=\sqrt{2\mu_0 P_0/(lwc)}$.}
We remark that a stronger magnomechanical coupling strength yields a greater microwave-optics entanglement but also a narrower parameter regime of the system being stable. For example, under the parameters of Figure 2b and the optimal detunings, the coupling can be as high as $G_m/2\pi=4.6$ MHz while keeping the system stable, which yields $E_{ca}\simeq 0.4$. 
%{\bf The bandwidth of the entangled fields is determined by the effective mechanical linewidth, which is about $\gamma_b^{\rm eff}  \simeq xx$ kHz, and is compatible with current superconducting qubit technology~\cite{JiePRX}. }

Having determined the optimal working conditions of the OMM system for the entanglement (i.e., $\tilde{\Delta}_c \simeq - \tilde{\Delta}_m \simeq  \omega_b$, and $\omega_b \gg \kappa_{m,c}$ for resolved sidebands), we optimize the entanglement over two parameters of the microwave cavity, i.e., the cavity-magnon detuning $\Delta_{am} \equiv \Delta_a-\tilde{\Delta}_m$ and the cavity decay rate $\kappa_a$, as shown in Figure 3a. Clearly, the resonant case $\Delta_{am}=0$ maximizes the microwave-optics entanglement, indicating that the entanglement is transferred from the magnomechanical subsystem (explicitly, the magnon-phonon parametric down-conversion process) via the microwave-magnon and optomechanical beamsplitter couplings (c.f. Figure 1b). It also shows that small cavity loss is always helpful as long as the stability is maintained. Figure 3b studies the effect of the temperature and mechanical damping on the optical entanglement. The entanglement is robust against temperature: For moderate damping rates $\gamma_b/2\pi < 10^3$ Hz, the entanglement can survive up to $\sim0.3$ K, whereas for a significantly large damping rate $\gamma_b/2\pi = 10^5$ Hz (quality factor $Q_b=400$), the entanglement can still be present up to $\sim0.1$ K.

\section{Conclusion}
We have presented a feasible and efficient approach to generate stationary microwave-optics entanglement in a microwave cavity-OMM system. The entanglement is achieved by exploiting dispersive magnetostriction and radiation pressure interactions. Specifically, by realizing the effective magnomechanical parametric down-conversion and optomechanical beamsplitter interactions, magnon-phonon entanglement is created and the entanglement is further distributed to optical and microwave cavity photons via the optomechanical and electromagnonic state-swap interactions, yielding stationary microwave-optics entanglement. Optimal working conditions are analyzed and the entanglement is robust against temperature even for significantly large mechanical damping. Such microwave-optics entanglement will find many distinctive applications in quantum networks and quantum information processing with hybrid quantum systems.

% Acknowledgements
\medskip
\section*{\textbf{Acknowledgements}} \par %delete if not applicable))
This work is supported by National Key Research and Development Program of China (Grant No. 2022YFA1405200) and National Natural Science Foundation of China (Nos. 92265202).

% References
\medskip

% Use the following code if you wish to generate your bibliography with BibTeX;
% replace the string "MSP-template" below with the name(s) of
% the BibTeX data base(s) you want to use.
% The resulting bibliography-output (the content of the .bbl file)
% must be pasted back into this file before submission.
% Please also include your BibTeX data base file(s) in your submission
% so that we can re-run BibTeX if necessary.
%
%\bibliographystyle{MSP}
%\bibliography{MSP-template}

\begin{thebibliography}{99}
	
	\bibitem{Wootters}
	C. H. Bennett, G. Brassard, C. Cr\'epeau, R. Jozsa, A. Peres, W. K. Wootters, Phys. Rev. Lett. {\bf 1993}, 70, 1895.
	\bibitem{Bouwmeester}
	D. Bouwmeester, J.-W. Pan, K. Mattle, M. Eibl, H. Weinfurter, A. Zeilinger, Nature {\bf 1997}, 390, 575.
	\bibitem{Furusawa}
	A. Furusawa, J. L. S\o rensen, S. L. Braunstein, C. A. Fuchs, H. J. Kimble, E. S. Polzik, Science {\bf 1998}, 282, 706.
	\bibitem{Kimble}
	H. J. Kimble, Nature {\bf 2008}, 453, 1023.
	\bibitem{Wehner}
	S. Wehner, D. Elkouss, R. Hanson, Science {\bf 2018}, 362, 303.
	\bibitem{Milburn}
	E. Knill, R. Laflamme, G. J. Milburn, Nature {\bf 2001}, 409, 46.
	\bibitem{VG}
	V. Giovannetti, S. Lloyd, L. Maccone, Nat. Photon. {\bf 2011}, 5, 222.
	\bibitem{Hensen}
	B. Hensen, H. Bernien, A. E. Dr\'eau, A. Reiserer, N. Kalb, M. S. Blok, J. Ruitenberg, R. F. L. Vermeulen, R. N. Schouten, C. Abell\'an, W. Amaya, V. Pruneri, M. W. Mitchell, M. Markham, D. J. Twitchen, D. Elkouss, S. Wehner, T. H. Taminiau, R. Hanson, Nature {\bf 2015}, 526, 682.
	\bibitem{Giustina}
	M. Giustina, M. A. M. Versteegh, S. Wengerowsky, J. Handsteiner, A. Hochrainer, K. Phelan, F. Steinlechner, J. Kofler, J.-\AA. Larsson, C. Abell\'an, W. Amaya, V. Pruneri, M. W. Mitchell, J. Beyer, T. Gerrits, A. E. Lita, L. K. Shalm, S. W. Nam, T. Scheidl, R. Ursin, B. Wittmann, A. Zeilinger, Phys. Rev. Lett. {\bf 2015}, 115, 250401.
	\bibitem{Shalm}
	L. K. Shalm, E. M.-Scott, B. G. Christensen, P. Bierhorst, M. A. Wayne, M. J. Stevens, T. Gerrits, S. Glancy, D. R. Hamel, M. S. Allman, K. J. Coakley, S. D. Dyer, C. Hodge, A. E. Lita, V. B. Verma, C. Lambrocco, E. Tortorici, A. L. Migdall, Y. Zhang, D. R. Kumor, W. H. Farr, F. Marsili, M. D. Shaw, J. A. Stern, C. Abell\'an, W. Amaya, V. Pruneri, T. Jennewein, M. W. Mitchell, P. G. Kwiat, J. C. Bienfang, R. P. Mirin, E. Knill, S. W. Nam, Phys. Rev. Lett. {\bf 2015}, 115, 250402.
	
	
	\bibitem{PK} 
	Z. Y. Ou, S. F. Pereira, H. J. Kimble, K. C. Peng, Phys. Rev. Lett. {\bf 1992}, 68, 3663.
	\bibitem{Kwiat} 
	P. G. Kwiat, K. Mattle, H. Weinfurter, A. Zeilinger, A. V. Sergienko, Y. Shih, Phys. Rev. Lett. {\bf 1995}, 75, 4337.
	\bibitem{fiber1} 
	J. E. Sharping, M. Fiorentino, P. Kumar, Opt. Lett. {\bf 2001}, 26, 367. 
	\bibitem{fiber2} 	
	X. Li, P. L. Voss, J. E. Sharping, P. Kumar, Phys. Rev. Lett. {\bf 2005}, 94, 053601.
	\bibitem{atom1}
	A. M. Marino, R. C. Pooser, V. Boyer, P. D. Lett, Nature {\bf 2009}, 457, 859.
	\bibitem{atom2}	
	Z. Qin, L. Cao, H. Wang, A. Marino, W. Zhang, J. Jing, Phys. Rev. Lett. {\bf 2014}, 113, 023602.
	\bibitem{dot1}
	O. Benson, C. Santori, M. Pelton, Y. Yamamoto, Phys. Rev. Lett. {\bf 2000}, 84, 2513.
	\bibitem{dot2}
	R. M. Stevenson, R. J. Young, P. Atkinson, K. Cooper, D. A. Ritchie, A. J. Shields, Nature {\bf 2006}, 439, 179.
	\bibitem{Gisin}
	S. Tanzilli, H. De Riedmatten, W. Tittel, H. Zbinden, P. Baldi, M. De Micheli, D. B. Ostrowsky, N. Gisin, Electron. Lett. {\bf 2001}, 37, 26.
	
	
	\bibitem{JPA1}
	C. Eichler, D. Bozyigit, C. Lang, M. Baur, L. Steffen, J. M. Fink, S. Filipp, A. Wallraff, Phys. Rev. Lett. {\bf 2011}, 107, 113601.	
	\bibitem{JPA2}
	N. Bergeal, F. Schackert, L. Frunzio, M. H. Devoret, Phys. Rev. Lett. {\bf 2012}, 108, 123902.
	\bibitem{JPA3}
	E. Flurin, N. Roch, F. Mallet, M. H. Devoret, B. Huard, Phys. Rev. Lett. {\bf 2012}, 109, 183901.
	\bibitem{DV2007}
	S. Pielawa, G. Morigi, D. Vitali, L. Davidovich, Phys. Rev. Lett. {\bf 2007}, 98, 240401.
	
	\bibitem{Menzel}
	E. P. Menzel, R. Di Candia, F. Deppe, P. Eder, L. Zhong, M. Ihmig, M. Haeberlein, A. Baust, E. Hoffmann, D. Ballester, K. Inomata, T. Yamamoto, Y. Nakamura, E. Solano, A. Marx, R. Gross, Phys. Rev. Lett. {\bf 2012}, 109, 250502.
	
	\bibitem{Tombesi}
	%V. Giovannetti, S. Mancini, and P. Tombesi, Europhys. Lett. {\bf 54}, 559 (2001); 
	S. Barzanjeh, D. Vitali, P. Tombesi, G. J. Milburn, Phys. Rev. A {\bf 2011}, 84, 042342. 
	\bibitem{LTian}
	%V. Giovannetti, S. Mancini, and P. Tombesi, Europhys. Lett. {\bf 54}, 559 (2001); 
	L. Tian, Phys. Rev. Lett. {\bf 2013}, 110, 233602. 
	\bibitem{YDWang}
	%V. Giovannetti, S. Mancini, and P. Tombesi, Europhys. Lett. {\bf 54}, 559 (2001); 
	Y.-D. Wang, A. A. Clerk, Phys. Rev. Lett. {\bf 2013}, 110, 253601.
	\bibitem{HXTang}
	C. Zhong, Z. Wang, C. Zou, M. Zhang, X. Han, W. Fu, M. Xu, S. Shankar, M. H. Devoret, H. X. Tang, L. Jiang, Phys. Rev. Lett. {\bf 2020}, 124, 010511. 
	
	\bibitem{SB1}
	S. Barzanjeh, E. S. Redchenko, M. Peruzzo, M. Wulf, D. P. Lewis, G. Arnold, J. M. Fink, Nature {\bf 2019}, 570, 480.
	\bibitem{SB2}
	J. Chen, M. Rossi, D. Mason, A. Schliesser, Nat. Commun. {\bf 2020}, 11, 943.
	
	\bibitem{Hybrid1}
	Z.-L. Xiang, S. Ashhab, J. Q. You, F. Nori, Rev. Mod. Phys. {\bf 2013}, 85, 623.
	\bibitem{Hybrid2}
	G. Kurizki, P. Bertet, Y. Kubo, K. Molmer, D. Petrosyan, P. Rabl, J. Schmiedmayer, Proc. Natl. Acad. Sci. USA {\bf 2015}, 112, 3866.	
	\bibitem{Hybrid3}
	A. A. Clerk, K. W. Lehnert, P. Bertet, J. R. Petta, Y. Nakamura, Nat. Phys. {\bf 2020}, 16, 257.	
	
	\bibitem{JiePRX1}
	R. J. Schoelkopf, S. M. Girvin, Nature {\bf 2008}, 451, 664. 
	\bibitem{JiePRX2}
	S. Pirandola, S. L. Braunstein, Nature {\bf 2016}, 532, 169. 
	\bibitem{JiePRX3}
	S. Krastanov, H. Raniwala, J. Holzgrafe, K. Jacobs, M. Lončar, M. J. Reagor, D. R. Englund, Phys. Rev. Lett. {\bf 2021}, 127, 040503. 
	\bibitem{JiePRX4}
	J. Li, Y.-P. Wang, W.-J. Wu, S.-Y. Zhu, J. Q. You, PRX Quantum {\bf 2021}, 2, 040344. 
	\bibitem{JiePRX5}
	J. Agust\'i, Y. Minoguchi, J. M. Fink, P. Rabl, Phys. Rev. A {\bf 2022}, 105, 062454. 
		
	\bibitem{Forsch1}
	T. Bagci, A. Simonsen, S. Schmid, L. G. Villanueva, E. Zeuthen, J. Appel, J. M. Taylor, A. S\o rensen, K. Usami, A. Schliesser, E. S. Polzik, Nature {\bf 2014}, 507, 81. %N. Malossi {\it et al.}, Phys. Rev. A {\bf 103}, 033516 (2021).
	\bibitem{Forsch2}
	R. W. Andrews, R. W. Peterson, T. P. Purdy, K. Cicak, R. W. Simmonds, C. A. Regal, K. W. Lehnert, Nat. Phys. {\bf 2014}, 10, 321. %N. Malossi {\it et al.}, Phys. Rev. A {\bf 103}, 033516 (2021).
	\bibitem{Forsch3}
	M. Forsch, R. Stockill, A. Wallucks, I. Marinkovi\'c, C. G\"artner, R. A. Norte, F. van Otten, A. Fiore, K. Srinivasan, S. Gr\"oblacher, Nat. Phys. {\bf 2020}, 16, 69. %N. Malossi {\it et al.}, Phys. Rev. A {\bf 103}, 033516 (2021).
	
	\bibitem{Tsang}
	M. Tsang, Phys. Rev. A {\bf 2010}, 81, 063837.
	\bibitem{Fink19}
	A. Rueda, W. Hease, S. Barzanjeh, J. M. Fink, npj Quantum Inf. {\bf 2019}, 5, 108.	
	\bibitem{Salamnogli}
	A. Salmanogli, D. Gokcen, H. Selcuk Gecim, Phys. Rev. Appl. {\bf 2019}, 11, 024075.	
	%%%%%%%%%%%%%%magnons
	


	\bibitem{SA}
	X. Zhang, C.-L. Zou, L. Jiang, H. X. Tang, Sci. Adv. {\bf 2016}, 2, e1501286.
	\bibitem{Jie18}
	J. Li, S.-Y. Zhu, G. S. Agarwal, Phys. Rev. Lett. {\bf 2018}, 121, 203601.
	
	\bibitem{JieL20}
	M. Yu, H. Shen, J. Li, Phys. Rev. Lett. {\bf 2020}, 124, 213604.
	
	\bibitem{Davis}
	C. A. Potts, E. Varga, V. Bittencourt, S. V. Kusminskiy, J. P. Davis, Phys. Rev. X {\bf 2021}, 11, 031053.
	\bibitem{Jie22}
	R.-C. Shen, J. Li, Z.-Y. Fan, Y.-P. Wang, J. Q. You, Phys. Rev. Lett. {\bf 2022}, 129, 123601.
	
	\bibitem{Fan}
	Z.-Y. Fan, R.-C. Shen, Y.-P. Wang, J. Li, J. Q. You. Phys. Rev. A {\bf 2022}, 105, 033507.
	
	
	\bibitem{S1}
	H. Huebl, C. W. Zollitsch, J. Lotze, F. Hocke, M. Greifenstein, A. Marx, R. Gross, S. T. B. Goennenwein, Phys. Rev. Lett. {\bf 2013}, 111, 127003. 
	\bibitem{S2}
	Y. Tabuchi, S. Ishino, T. Ishikawa, R. Yamazaki, K. Usami, Y. Nakamura, Phys. Rev. Lett. {\bf 2014}, 113, 083603. 
	\bibitem{S3}
	X. Zhang, C.-L. Zou, L. Jiang, H. X. Tang, Phys. Rev. Lett. {\bf 2014}, 113, 156401. 
	
	
	%\bibitem{Hill}A. V. Chumak, V. I. Vasyuchka, A. A. Serga, and B. Hillebrands, Nat. Phys. {\bf 11}, 453 (2015).
	%\bibitem{Naka19}D. Lachance-Quirion, Y. Tabuchi, A. Gloppe, K. Usami, and Y. Nakamura, Applied Physics Express {\bf 12}, 070101 (2019).
	%\bibitem{Yuan}H. Y. Yuan, Y. Cao, A. Kamra, R. A. Duine, and P. Yan, Phys. Rep. {\bf 965}, 1 (2022).
	
	
	
	%\bibitem{Jie19}J. Li and S.-Y. Zhu, New J. Phys. {\bf 21}, 085001 (2019).
	
	%\bibitem{Jie19a}J. Li, S. Y. Zhu, and G. S. Agarwal, Phys. Rev. A {\bf 99}, 021801(R) (2019).
	
	%\bibitem{QST}J. Li and S. Gr\"oblacher, Quantum Sci. Technol. {\bf 6}, 024005 (2021).
	
	
	
\bibitem{APE}
D. Lachance-Quirion, Y. Tabuchi, A. Gloppe, K. Usami, and Y. Nakamura, Appl. Phys. Express {\bf 2019}, 12, 070101.

\bibitem{x1}
Z.-X. Liu and H. Xiong, Opt. Lett. {\bf 2020}, 45, 5452.
\bibitem{x2}
Z.-X. Liu and Y.-Q. Li, Photonics Research {\bf 2022}, 10, 2786.
\bibitem{x3}
Z.-X. Liu, H. Xiong, and Y. Wu, Phys. Rev. B {\bf 2019}, 100, 134421.
\bibitem{x4}
J. Xie, S. Ma, and F. Li, Phys. Rev. A {\bf 2020}, 101, 042331.
	
	
	
	\bibitem{Kittel}
	C. Kittel, Phys. Rev. {\bf 1948}, 73, 155.
	
	\bibitem{MA}
	M. Aspelmeyer, T. J. Kippenberg, F. Marquardt, Rev. Mod. Phys. {\bf 2014}, 86, 1391.
	
	\bibitem{QST23}
	Z. Fan, H. Qian and J. Li, Quantum Sci. Technol. {\bf 2023}, 8, 015014.
	
	
	\bibitem{PRAp}
	F. Heyroth, C. Hauser, P. Trempler, P. Geyer, F. Syrowatka, R. Dreyer, S. G. Ebbinghaus, G. Woltersdorf, G. Schmidt, Phys. Rev. Appl. {\bf 2019}, 12, 054031.
	
	\bibitem{SG}
	S. Gr\"oblacher, K. Hammerer, M. R. Vanner, M. Aspelmeyer, Nature {\bf 2009}, 460, 724.
	
	\bibitem{Harris}
	J. D. Thompson, B. M. Zwickl, A. M. Jayich, F. Marquardt, S. M. Girvin, J. G. E. Harris, Nature {\bf 2008}, 452, 72.
	
	
	%\bibitem{note}
	%We, however, do not use thin YIG films, where high-frequency (GHz) phonon modes couple linearly (and strongly) to the magnon mode~\cite{KAn1,KAn2}. The linear coupling is not desired for our purpose of producing entanglement.
	\bibitem{KAn1}
	F. Godejohann, A. V. Scherbakov, S. M. Kukhtaruk, A. N. Poddubny, D. D. Yaremkevich, M. Wang, A. Nadzeyka, D. R. Yakovlev, A. W. Rushforth, A. V. Akimov, M. Bayer, Phys. Rev. B {\bf 2020}, 102, 144438.
	\bibitem{KAn2}
	K. An, A. N. Litvinenko, R. Kohno, A. A. Fuad, V. V. Naletov, L. Vila, U. Ebels, G. de Loubens, H. Hurdequint, N. Beaulieu, J. Ben Youssef, N. Vukadinovic, G. E. W. Bauer, A. N. Slavin, V. S. Tiberkevich, O. Klein, Phys. Rev. B {\bf 2020}, 101, 060407.	
	
	
	\bibitem{Dong}
	Z. Shen, G. Xu, M. Zhang, Y. Zhang, Y. Wang, C. Chai, C. Zou, G. Guo, and C. Dong, Phys. Rev. Lett. {\bf 2022} 129, 243601.
	
	\bibitem{Kac1}
	R. Benguria, M. Kac, Phys. Rev. Lett. {\bf 1981}, 46, 1.
	\bibitem{Kac2}
	V. Giovannetti, D. Vitali, Phys. Rev. A {\bf 2001}, 63, 023812.
	
	\bibitem{DV07}
	D. Vitali, S. Gigan, A. Ferreira, H. R. B\"ohm, P. Tombesi, A. Guerreiro, V. Vedral, A. Zeilinger, M. Aspelmeyer, Phys. Rev. Lett. {\bf 2007}, 98, 030405.
	
	\bibitem{LN1}
	G. Vidal, R. F. Werner, Phys. Rev. A {\bf 2002}, 65, 032314.
	\bibitem{LN2}
	G. Adesso, A. Serafini, F. Illuminati, Phys. Rev. A {\bf 2004}, 70, 022318.
	\bibitem{LN3}
	M. B. Plenio, Phys. Rev. Lett. {\bf 2005}, 95, 090503.
	
	\bibitem{Genes}
	C. Genes, A. Mari, P. Tombesi, D. Vitali, Phys. Rev. A {\bf 2008}, 78, 032316.
	
	\bibitem{Gut}
	C. Gut, K. Winkler, J. Hoelscher-Obermaier, S. G. Hofer, R. Moghadas Nia, N. Walk, A. Steffens, J. Eisert, W. Wieczorek, J. A. Slater, M. Aspelmeyer, K. Hammerer, Phys. Rev. Res. {\bf 2020}, 2, 033244.
	
	
	\bibitem{Jie19a1}
	M. Yu, S.-Y. Zhu, J. Li, J. Phys. B {\bf 2020}, 53, 065402.
	\bibitem{Jie19a2}
	J. Li, S. Gr\"oblacher, Quantum Sci. Technol. {\bf 2021}, 6, 024005.	
	
	%\bibitem{note2}
	%The time average of energy density is $\rho_e=H_d^2/(2\mu_0)$, with $\mu_0$ being the vacuum magnetic permeability. Therefore, the microwave drive power $P_0=\rho_e Ac$, where $A=l \times w$ is the cross-sectional area of the YIG cuboid with $l$ and $w$ being its length and width, and $c$ is the speed of the electromagnetic wave propagating in vacuum. The drive magnetic field is applied in the direction perpendicular to the cross-section. This then yields the relation between the drive magnetic field $H_d$ and the power $P_0$, i.e., $H_d=\sqrt{2\mu_0 P_0/(lwc)}$.
	
	
\end{thebibliography}

% Figures/tables and captions
% Permission statements are required for all figures reproduced or adapted from previously published articles/sources. Please also ensure that all necessary permissions to reproduce images have been received
% Please remove these statements for original figures

% Please provide Biographies and photos for Essays, Feature Articles, Progress Reports, Reviews, and Perspectives for those authors who should be highlighted  
% These should be at most 100 words long
% For other article types this section can be removed
% Photographs should be 40mm broad and 50 mm high

 \begin{figure}[h]
	\centering
	\includegraphics[width=0.55\linewidth]{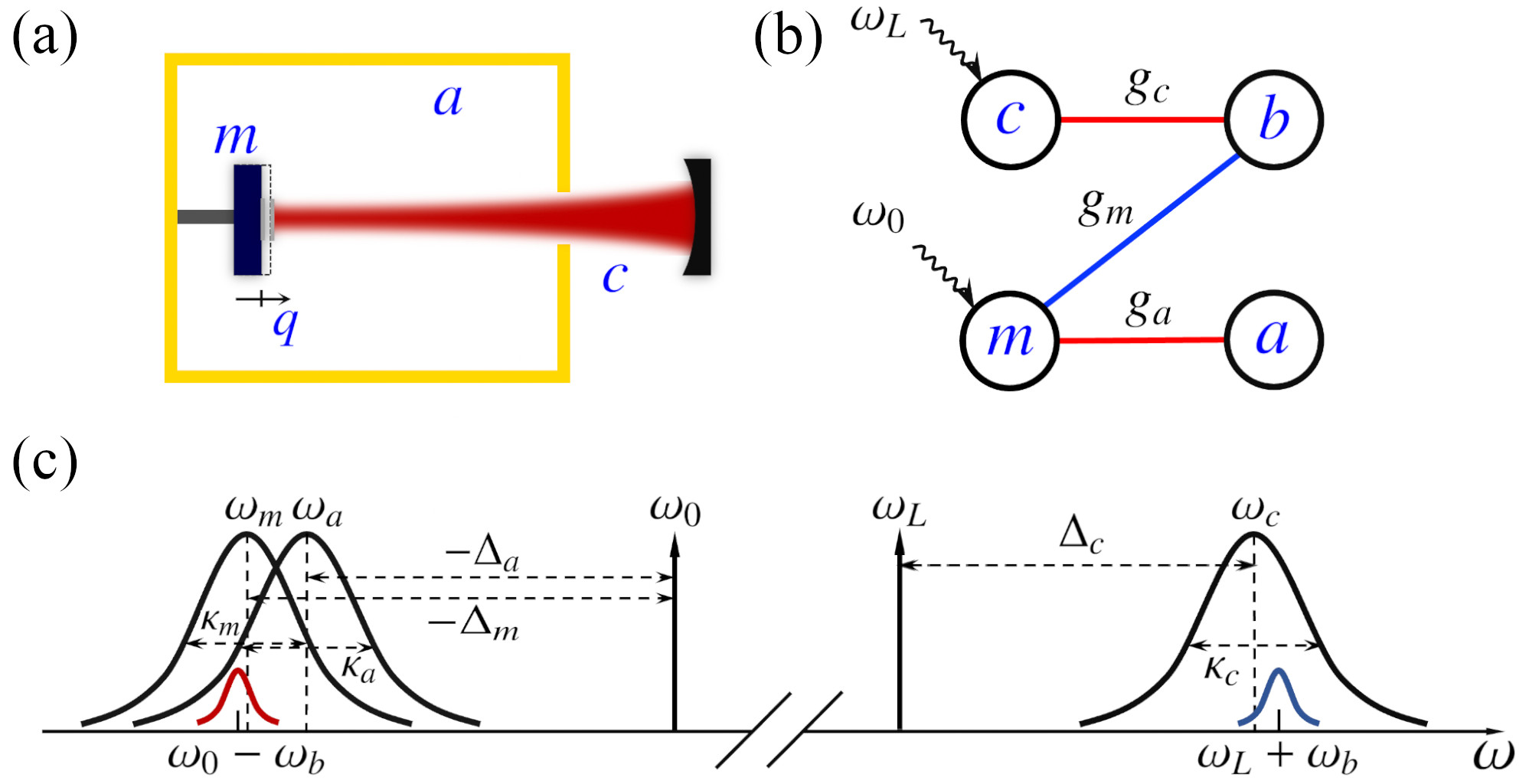}
	\caption*{Figure 1. a,b) Sketch of the cavity-OMM system. A magnon mode $m$ in a YIG crystal couples to a microwave cavity mode $a$ and to an optical cavity mode $c$ via the mechanical vibration $b$ induced by the magnetostriction. The blue (red) line in (b) denotes the effective parametric down-conversion (beamsplitter) interaction. c) Mode frequencies and linewidths used in the protocol. When the optical cavity is resonant with the (blue) anti-Stokes sideband of the driving laser at frequency $\omega_L+\omega_b$, and the magnon and microwave cavity modes are resonant with the (red) Stokes sideband of the microwave drive field at frequency $\omega_0-\omega_b$, and when the mechanical sidebands are resolved ($\omega_b \gg \kappa_j$), a stationary microwave-optics entangled state is generated.}
	\label{fig1}
\end{figure}

\begin{figure}[h]
	\centering
	\includegraphics[width=0.57\linewidth]{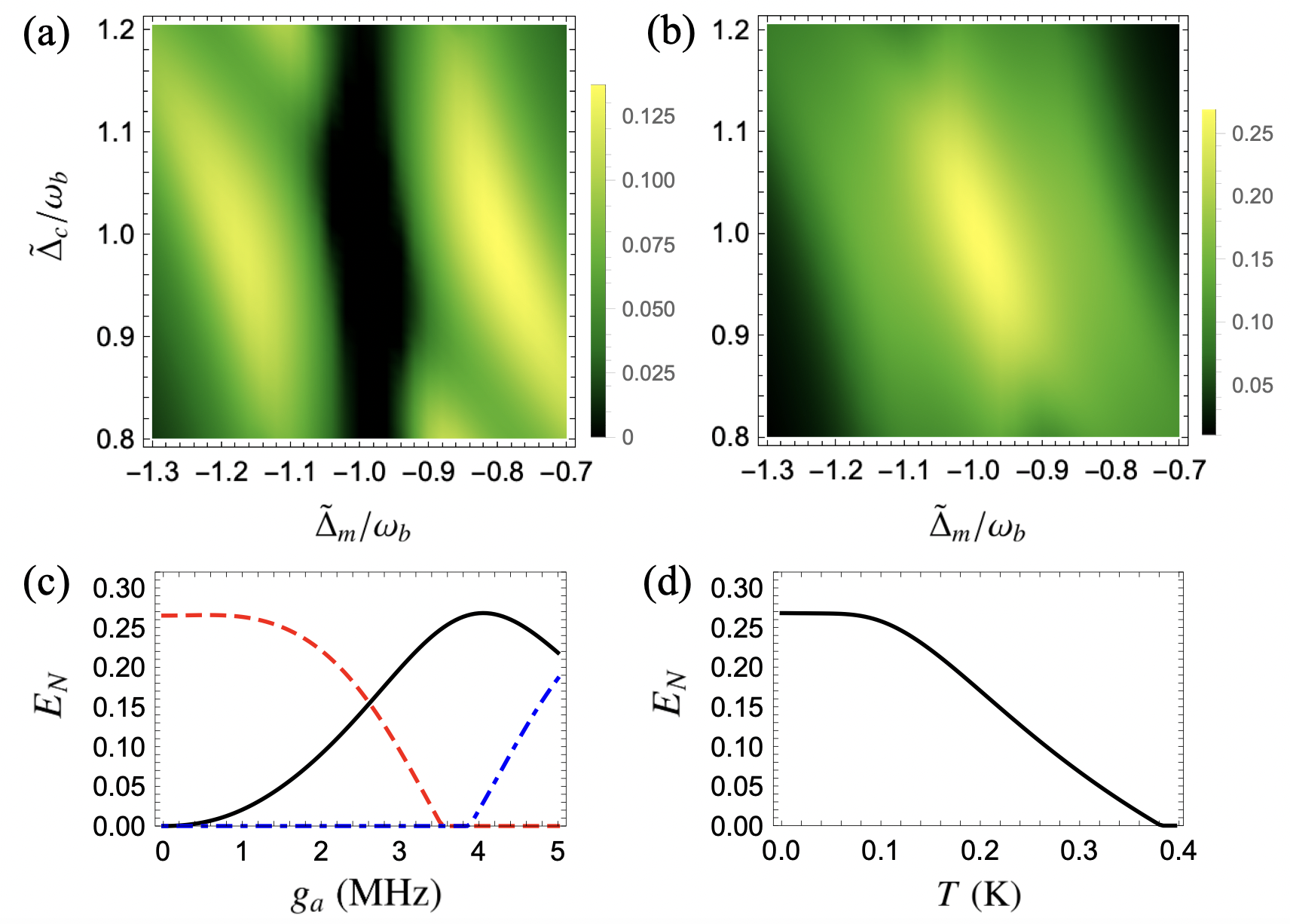}
	\caption*{Figure 2. a,b) Density plot of the steady-state magnon-optics entanglement $E_{cm}$ and microwave-optics entanglement $E_{ca}$ versus effective detunings $\tilde{\Delta}_m$ and $\tilde{\Delta}_c$. c) $E_{cm}$ (red dashed), $E_{ca}$ (black solid), and microwave-mechanics entanglement $E_{ab}$ (blue dot-dashed) versus the microwave-magnon coupling rate $g_a$ at optimal detunings $\tilde{\Delta}_c= - \tilde{\Delta}_m=\omega_b$, and $\Delta_a=\tilde{\Delta}_m$. d) $E_{ca}$ versus temperature $T$ at the optimal detunings. $g_a/2\pi = 4$ MHz is used in (a), (b) and (d). See the text for the rest of the parameters.}
	\label{fig2}
\end{figure}

\begin{figure}[h]
	\centering
	\includegraphics[width=0.63\linewidth]{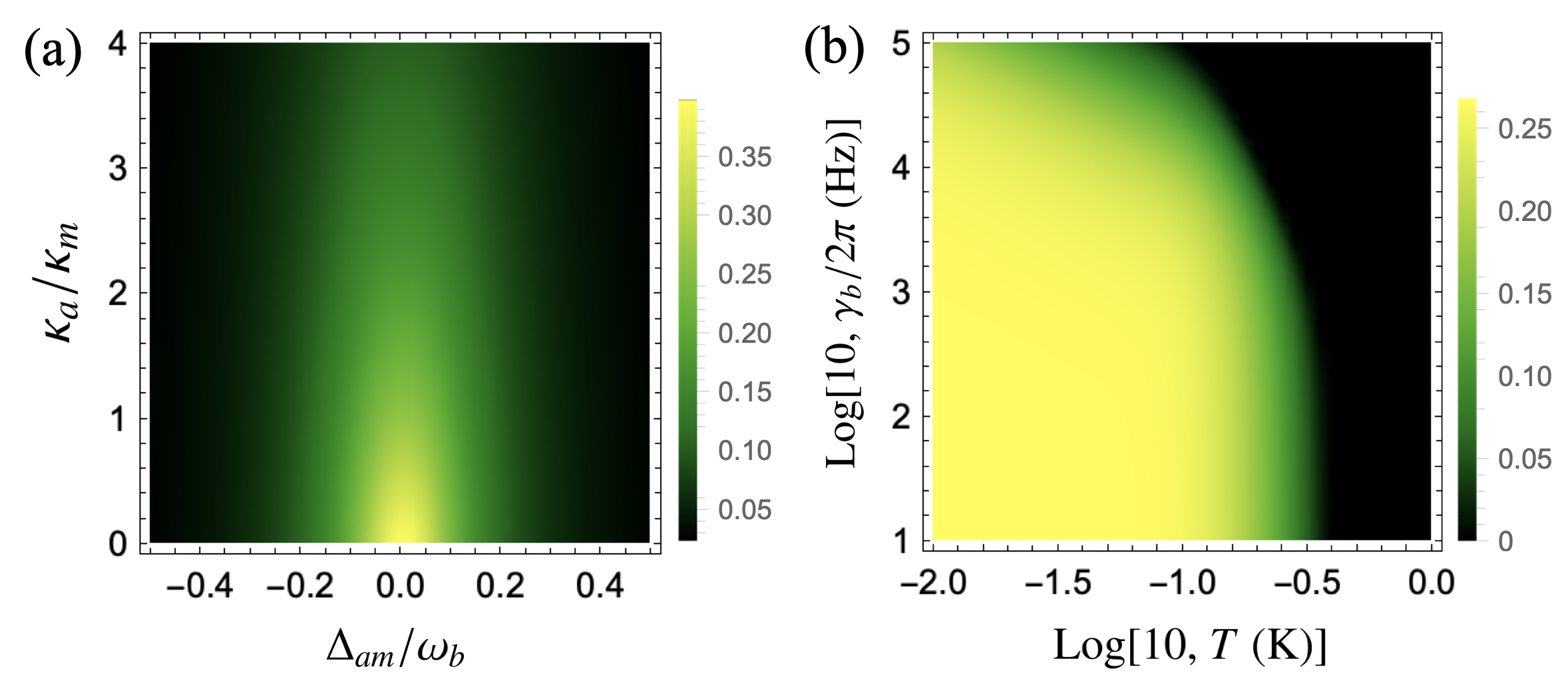}
	\caption*{Figure 3. Stationary microwave-optics entanglement $E_{ca}$ versus a) the cavity-magnon detuning $\Delta_{am}$ and the cavity decay rate $\kappa_a/\kappa_m$ (with fixed $\kappa_m$ and $ \tilde{\Delta}_m = - \tilde{\Delta}_c = -\omega_b$); b) temperature $T$ and mechanical damping rate $\gamma_b$. The other parameters are the same as in Figure 2b.}
	\label{fig3}
\end{figure}

%\section*{\textbf{Table of Contents}} \par %delete if not applicable))
%Text:

%A new mechanism to generate microwave-optics entanglement in a cavity optomagnomechanical system is proposed. Magnon-phonon entanglement is created via magnomechanical parametric down-conversion and further distributed to optical and microwave photons via simultaneous optomechanical beamsplitter interaction and electromagnonic state-swap interaction, yielding stationary microwave-optics entanglement. The microwave-optics entanglement finds particularly important applications in hybrid quantum networks and quantum information processing with hybrid systems.\\

%\noindent Figure:
 %\begin{figure}[h]
%	\includegraphics[width=0.4\linewidth]{ToCfigure.jpg}
%	\caption*{ToC Figure Entry}
%\end{figure}

\end{document}